# Metallic nanoparticle on micro ring resonator for bio optical detection and sensing

**Ali Haddadpour**[1,3] **and Yasha Yi**[2, 3, 4, *]

[1]*Faculty of Electrical and Computer Engineering, University of Tabriz, Tabriz, Iran*
[2] *New York University, New York, NY 10012, USA*
[3]*CUNY Graduate Center, New York, NY 10016, USA*
[4]*M.I.T., Cambridge, MA 02139, USA*
*yys@alum.mit.edu

**Abstract:** We have numerically investigated the unique effects of *metallic* nanoparticle on the ring resonator, especially multiple Au nanoparticles on the micro ring resonator with the 4-port configuration on chip. For the Au nanoparticle, because it has smaller real refractive index than air and large absorption refractive index, we found that there is a *blue* shift for the ring resonance wavelength, instead of *red* shift normally observed for dielectric nanoparticles. The drop port intensity is strongly dependent on both number and size of nanoparticles, while relatively independent on position of nanoparticles. The correlation between the penetration depth of Au and the resonance mode evanescent tail is also discussed to reveal the unique properties of Au nanoparticle to be used for detection, sensing and nano medicine.

---

## References and links

---

## 1. Introduction

Optical resonator has generated wide interests in the detection and sensing field. For the relatively high Q microresonators, a small change on the refractive index can be detected from the shift of resonance wavelength. Recently, the splitting of the resonance modes has been observed, which is caused as a consequence of clockwise and anti clockwise propagating mode coupling. This phenomenon has been proposed for various applications, such as photonic molecules [1-7]. Nanoparticles have been heavily used in the optical detection and sensor area, as fast, non-invasive, and potentially label-free techniques are becoming more important for bio-sensing, gas sensing, chemical sensing and nano medicine fields. For example, metal nanoparticles are used as contrast agents in bio molecule sensing. Semiconductor nanoparticles are used as single photon emitters in quantum information processing, and as fluorescent markers for biological processes. Nanoshells with special engineering methods are used for cancer therapies and photothermal tumor ablation. Polymer nanoparticles are employed as calibration standards and probes in biological imaging in functionalized form [8-11]. The synergy between microresonator and nanoparticle is becoming more important with the rapid progress of nanophotonics field.

The influence on micro resonators by *dielectric* nanoparticles have been intensively studied recently, fiber tip is used to study the resonance mode profile, especially from the splitting of resonance mode. The mean resonance mode wavelength shift, splitting bandwidth as well as their dependence on dielectric nanoparticle size and position have also been studied by many groups [12-13].

Optical sensor (including bio sensor, chemical sensor and gas sensor) based on Whispering Gallery Mode (WGM) micro resonator has generated worldwide interests in this emerging filed involving dielectric nanoparticles or dielectric bio layer [14-23]. However, the microsphere is made manually, which results in large scale manufacture of optical sensor devices very challenging. The integrated micro ring resonator doesn't have this limit as we can use semiconductor microelectronics process and millions of devices can be fabricated and integrated on a single chip with nano scale precision.

*Metallic* nanoparticles have also been used intensively on detection and sensing, especially in the form of Surface Plasmons Polariton (SPP), as the surface mode generated at the interface between the dielectric surface and metallic surface is strongly confined at the interface, which can be utilized for many potential applications [24]. Recently, in the conventional 4-port micro resonator configuration, it is found the metallic nanoparticles can be used as a strong scatters on the micro ring resonators, which induces large reflection signals at an output port, which is normally dark [25-26]. The interaction between *metallic* nanoparticles and micro ring resonators has revealed many interesting phenomena and hasn't been emphasized much in previous studies as *dielectric* nanoparticles.

## 2. On chip microring resonator device structure and simulation method

Here we have numerically demonstrated a unique result by Au nanoparticles, when it is adsorbed at the edge of micro ring resonator. Compared to the resonance position without any Au nanoparticle, it was found that there is a *blue* shift for the resonance peak, which is opposite to the resonance wavelength shift direction when the dielectric nanoparticles are adsorbed onto the micro ring resonator. Due to the unique refractive index properties of Au, the number and position effects are also appealing and investigated in detail in this paper.

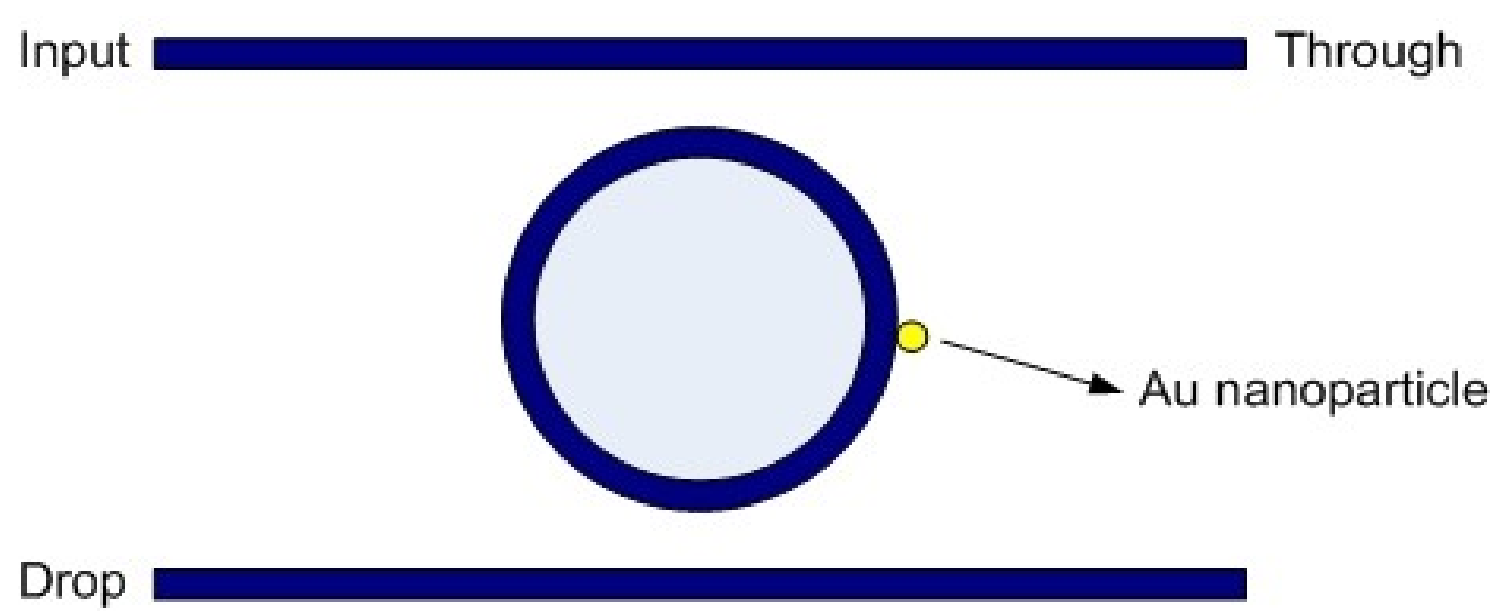

**Figure 1** The on-chip four port micro ring resonator configuration. The core of the micro ring resonator and the two bus waveguides is Si (n=3.48), with bottom cladding $SiO_2$ (n=1.45) and air cladding on top. The waveguide width is 200nm and satisfies the single mode condition. The Au nanoparticle is adsorbed on the micro ring resonator.

We used the conventional 4-port micro ring resonator configuration on a chip, as illustrated in Fig.1. The micro ring resonator is 4µm in diameter, and the ring waveguide width is 200nm as a single mode waveguide, the thickness of the ring resonator and bus waveguide is 250nm. The two bus waveguides are evanescently coupled to the micro ring resonator, with the coupling gap 100nm. In this work, we simulated the Si ring resonator and coupled waveguide system, with $SiO_2$ as the bottom cladding and air as the top cladding. The Au nanoparticle was placed at the outside edge of the micro ring resonator. The refractive index of Si is 3.48, $SiO_2$ is 1.45 and the dispersion relation of Au around wavelength 1.55µm is used [27].

We used Finite Difference Time Domain (FDTD) method in three dimensions to simulate the 4-port micro ring resonator with/without Au nanoparticles [28]. Due to the small size of nanoparticles, fine grid size as small as 2nm and sufficient long evolution time steps are used to check the reliability of the simulation until the optimized grid size and time steps were found to reduce the required memory and simulation time. Perfect Matched Layer (PML) absorbing boundary condition is used for the entire simulation window (10μmx10μm). The bus waveguide was excited with a Gaussian pulse which covers the wavelength window around 1.55μm, the detected signal at Drop port was Fourier transformed to obtain the Drop port vs. wavelength information.

## 3. Multiple Au nanoparticle effects on microring resonator and simulation results

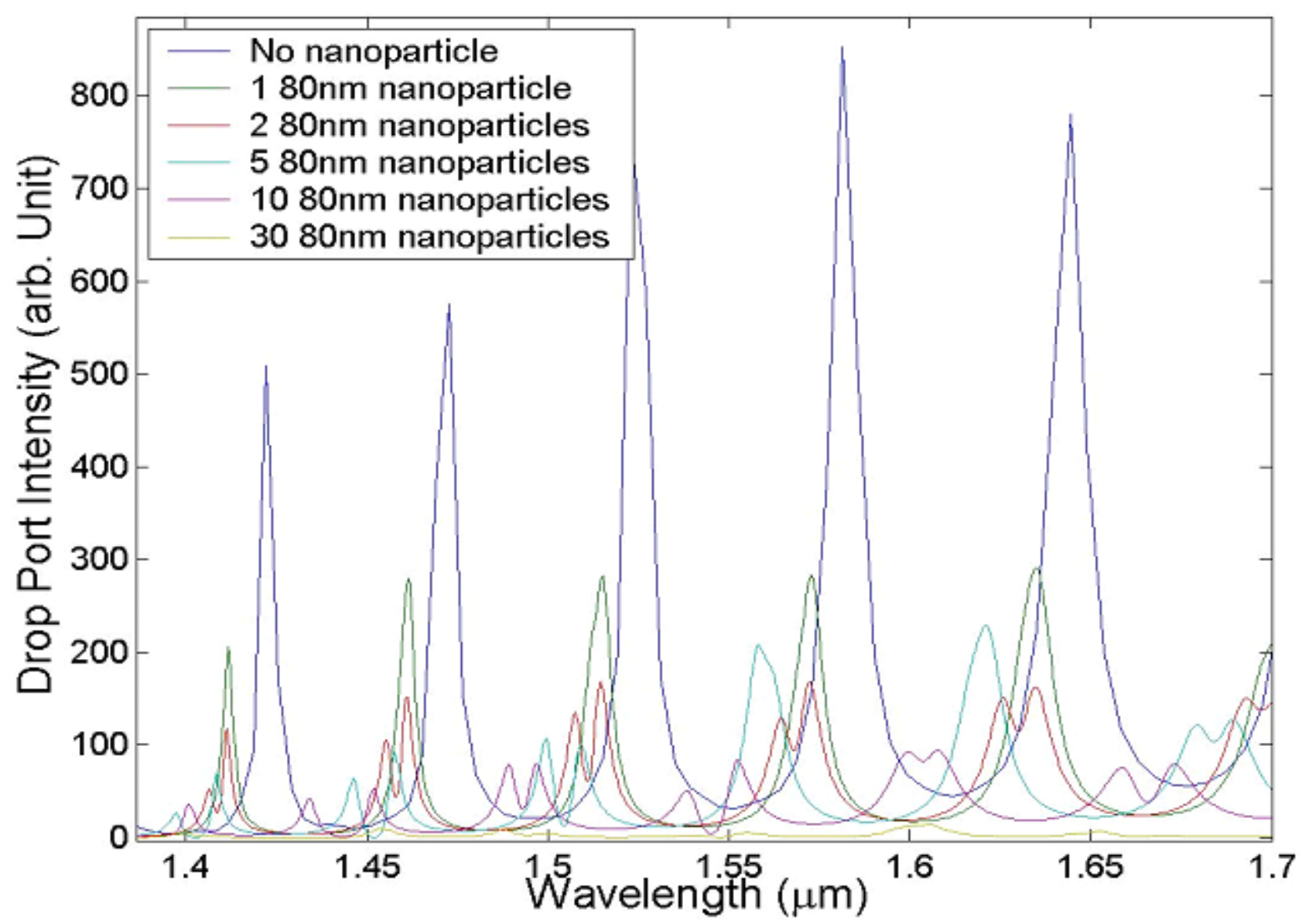

(a)

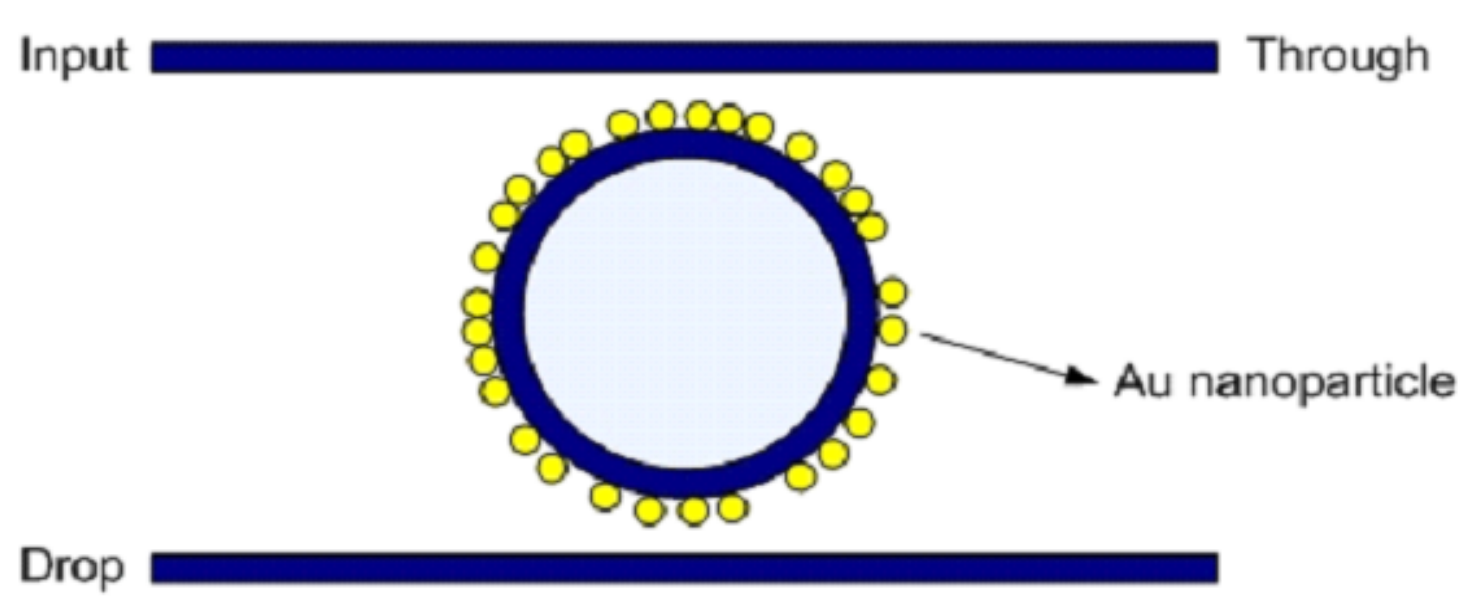

(b)

**Figure 2** (a) The Drop port intensity with different number of 80nm size Au nanoparticles adsorbed on the micro ring resonator. (b) the example of multiple Au nanoparticles randomly distributed on the micro ring resonator.

Similar to the dielectric nanoparticles, the metallic nanoparticles will make the resonance wavelength shift and broadening of the splitting bandwidth. As illustrated in Fig.2, the main difference of the Au nanoparticle is its smaller real refractive index than air and very large imaginary part (absorption part), which leads to the *blue* shift of the resonance mode wavelength position. As the nanoparticle numbers reach a certain number, the splitting of the resonance began to appear, within our simulation resolution. For the 80nm Au nanoparticle on the micro ring resonator with the nanoparticle number increasing from one to two, it is observed that both splitting modes are *blue* shifted. This phenomenon is unique as it provides us a very convenient approach to distinguish the dielectric nanoparticles and the Au nanoparticles, both are used extensively for sensing and nano medicine field. Furthermore, the intensity of Drop port is reduced rapidly with the increasing number of Au nanoparticles (thirty in this case), which represents its large absorption characteristics at this wavelength. For sensing applications using Au nanoparticles or other metallic nanoparticles, it is inferred from this work that there is a limit for the number of metallic nanoparticles adsorbed on the ring resonator, as there is normally large absorption for metallic nanoparticles. When the number reaches a certain point – critical number (30 Au nanoparticles in this case), the interaction between metallic nanoparticles and micro ring resonator is becoming so strong that they completely degrade the resonance – the Q is strongly degraded and the intensity at Drop port is approaching zero.

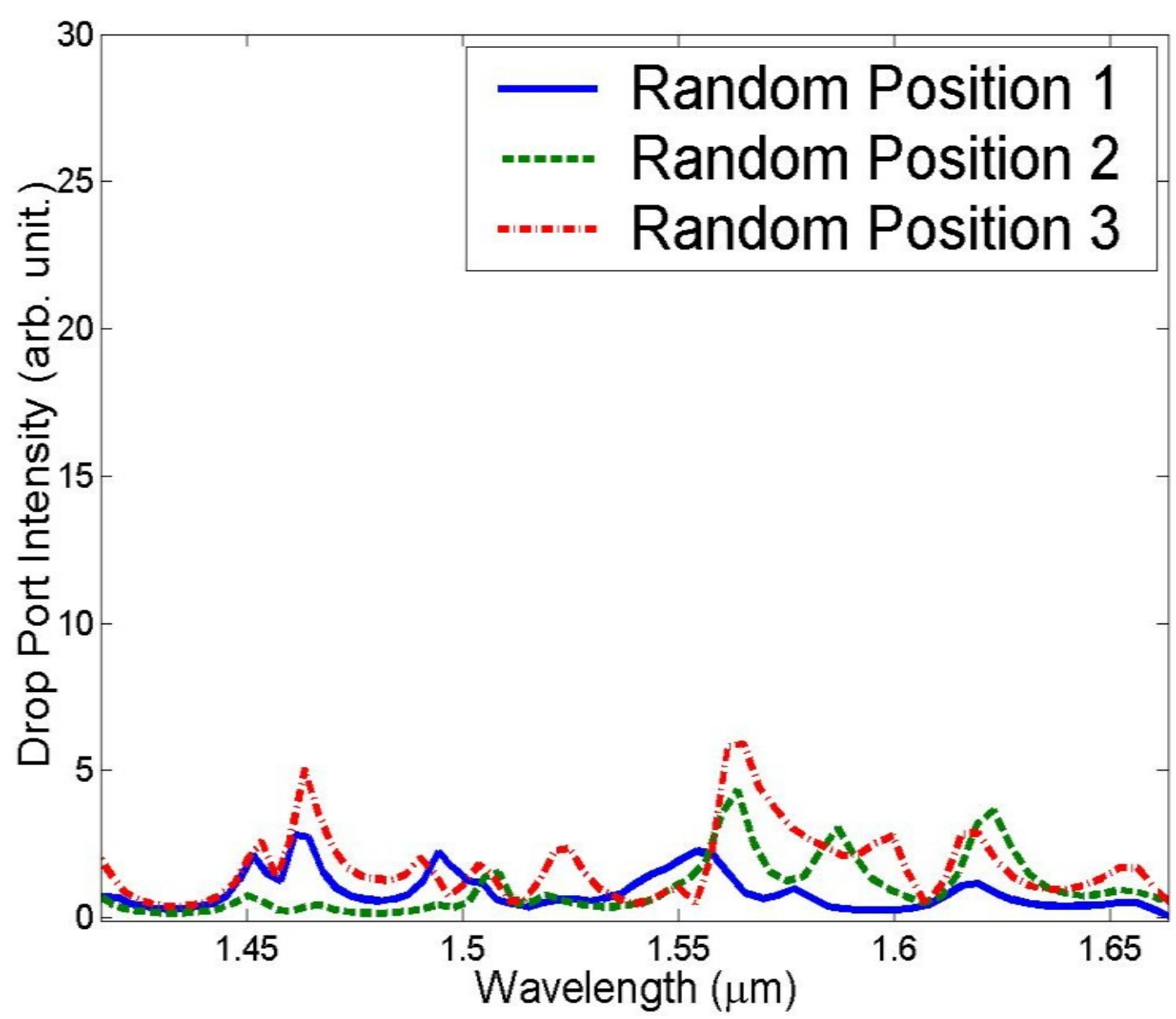

**Figure 3** The Drop port intensity with three random positions of 30 Au nanoparticles with 80nm in size.

For using Au nanoparticle and Drop port as a detection mechanism, the dependence on position is also important as the Au nanoparticles are possible to be adsorbed randomly on the micro ring resonator, the relatively position independence is necessary. To this purpose, for thirty Au naoparticles with 80nm in size, we have randomly distributed the 30 Au nanoparticles on the micro ring resonator and compared the Drop intensity in Fig.3. It is shown that the intensity at Drop port for three random positions is at the same order and this result demonstrates the relatively independence of the Au nanoparticle position on the micro ring resonator.

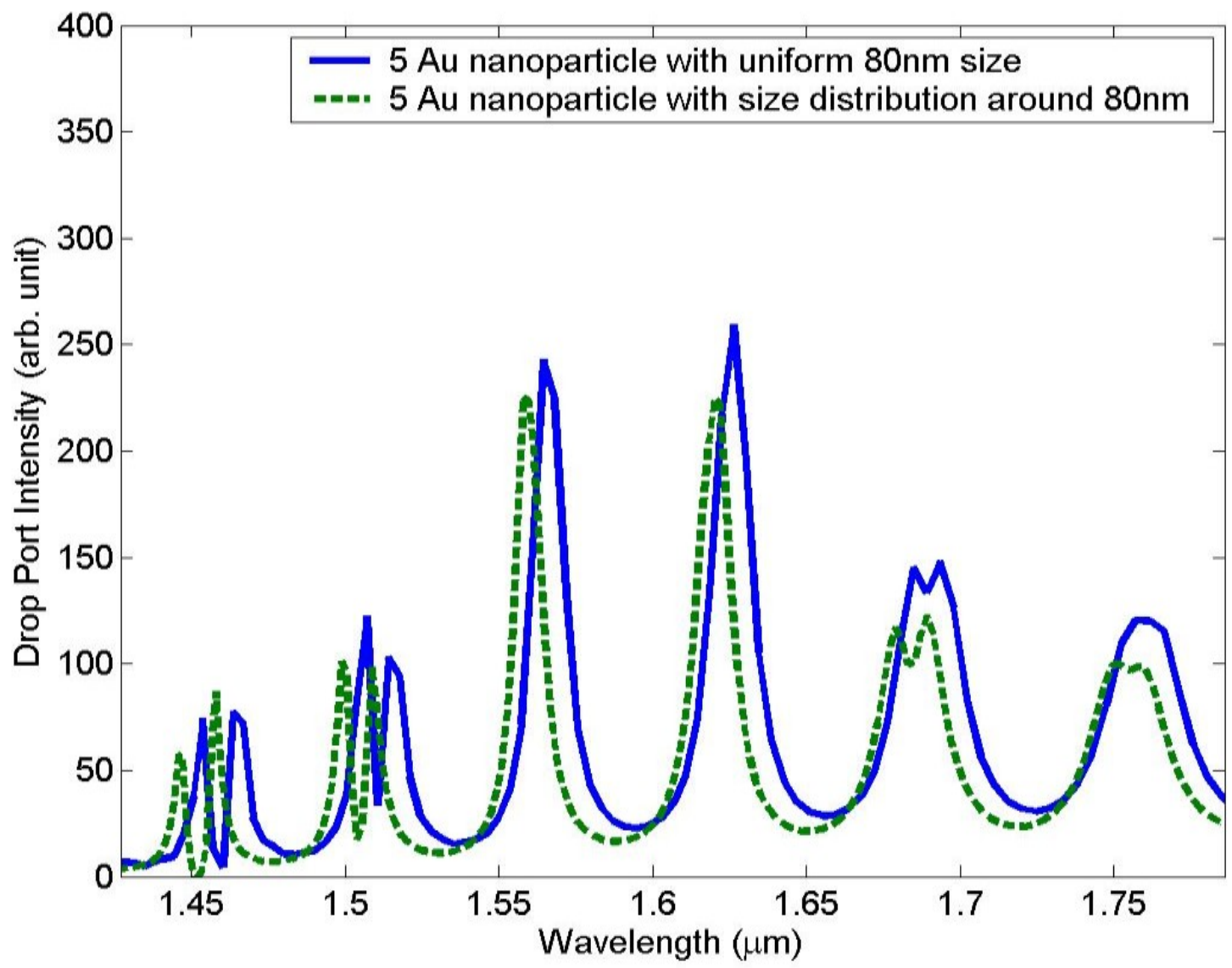

**Figure 4** The Drop port intensity with Au nanoparticle size distribution, 5 Au nanoparticles with uniform 80nm in diameter (solid line) and 5 Au nanoparticles with random size distribution around 80nm in size.

Metallic nanoparticle size uniformity is very important for practical sensing and detection, as the nanoparticle size normally has a distribution around the target nanoparticle size which we would like to use. To study the effect of uniformity of the size of the Au nanoparticles on the performance of the integrated micro ring resonator, for 5 Au nanoparticle case with 80nm size, we have randomly chosen the nanoparticle size which has certain distribution around 80nm. Fig.4 is the comparison between the uniform size nanoparticles and nanoparticles with certain distribution, the overall signal from drop port is almost the same although there is some small difference. The result demonstrates the robustness of our sensing mechanism using Au nanoparticles which can tolerate certain non uniformity of Au nanoparticles.

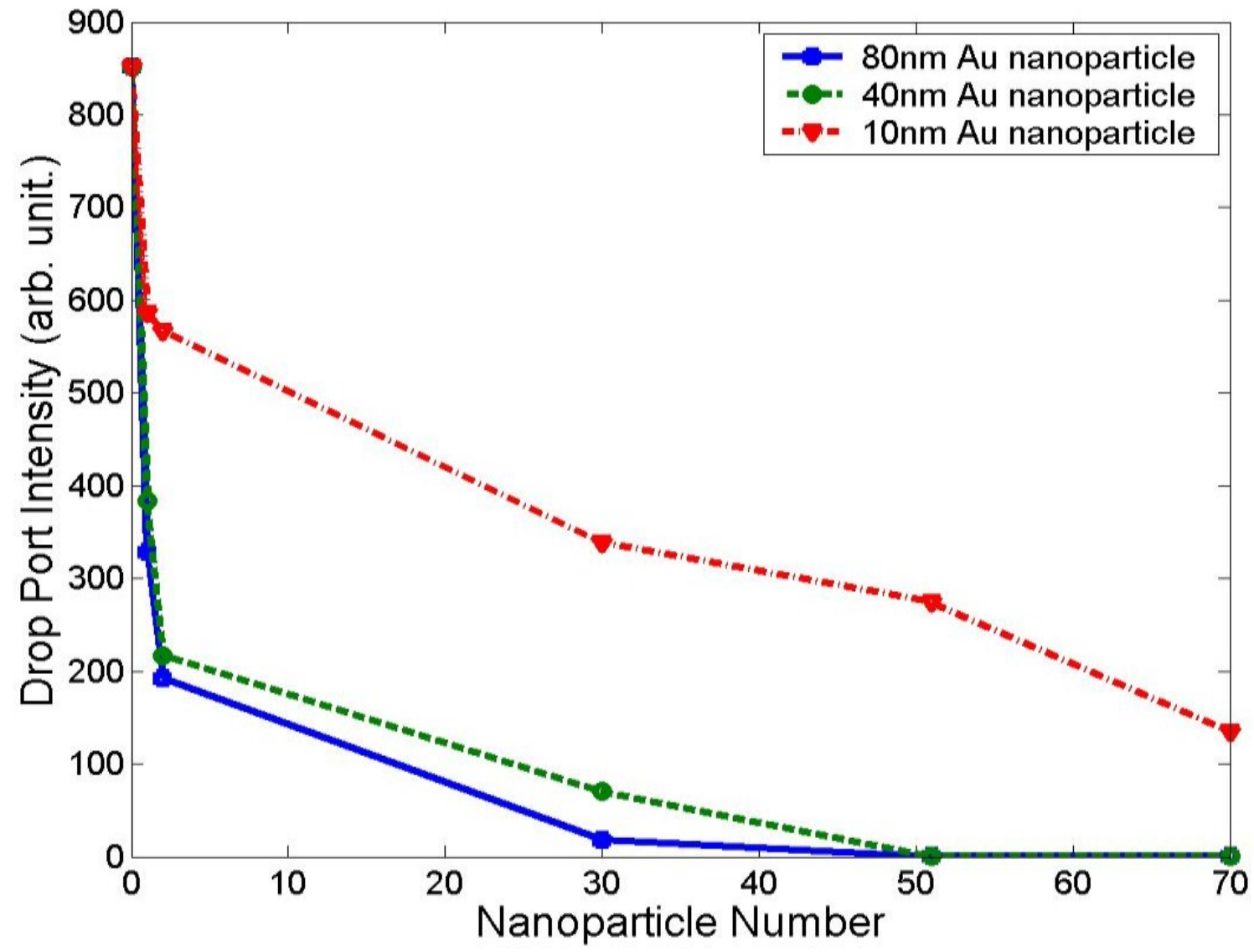

**Figure 5** The Drop intensity vs. nanoparticle number for different size of Au nanoparticles.

We also studied the dependence on Au naoparticle size, which is illustrated in Fig. 5. 10nm, 40nm and 80nm Au nanoparticle are compared at Drop port with the different number of Au nanoparticles. It is observed the intensity at Drop port for 10nm size Au nanoparticle is reduced in a much slower pace than that for the 40nm and 80nm Au nanoparticle. The reduction of the intensity is mainly caused by the strong absorption of Au nanoparticles, around 1.55μm; the penetration depth of Au is about 45nm. The dependence on nanoparticle size for the Drop port reveals the correlation between the penetration depths, nanoparticle size and resonance mode evanescent tail length, which might be utilized to measure the Au naoparticle size. For optical sensing and detection purpose, the optimized Au nanoparticle size should meet two requirements, one is relatively large shift when the nanoparticle is adsorbed to the ring resonator, the other is the slow Q degradation ratio when more Au naoparticles are adsorbed on the micro ring resonator. Based on the results in Fig.2 and Fig.5, the optimized size of Au nanoparticle is estimated as 40nm, which is around the penetration depth.

## 4. Conclusion

In summary, we have numerically demonstrated the unique optical response behavior for Au naoparticle in the 4-port micro ring resonator configuration, which can be utilized for single nanoparticle detection and related applications. The *blue* shift of the resonance mode position at Drop port due to the Au nanoparticle is drastically different from that of dielectric nanoparticles, which are widely used for various applications in detection, sensing and biomedical field. For sensing, the unwanted nanoparticle adsorbed on the micro ring resonator is mostly dielectric nanoparticles, which may be mixed with true signal, the unique *blue* shift by Au nanoparticle could be utilized to differentiate from the dielectric nanoparticles, therefore, it can be utilized to increase the signal to noise ratio. Due to large absorption of Au, the number of nanoparticles will reach a critical number before the micro ring resonator can still maintain an effective optical resonator device. The results on position and size dependence suggest the robustness to use Au nanoparticle for future applications in detection, sensing and nano medicine field.

## Acknowledgements

We thank the support from Microsystems Technology Laboratory and Center for Materials Science and Engineering at MIT and the High Performance Computing center at CUNY.